\documentclass[twocolumn,superscriptaddress,showpacs]{revtex4}
\usepackage{amssymb,epsfig}

\begin{document}
\title{
A Fitter Code for Deep Virtual Compton Scattering\\
and Generalized Parton Distributions.} 
\author{M. Guidal,\\
Institut de Physique Nucl\'eaire d'Orsay,\\
91405 Orsay, FRANCE 
}

\begin{abstract}
We have developed a fitting code based on the leading-twist handbag
Deep Virtual Compton Scattering (DVCS) amplitude in order to extract 
Generalized Parton Distribution (GPD) information from DVCS observables
in the valence region. 
In a first stage, with simulations and pseudo-data, we show that the full 
GPD information can be recovered from experimental data if enough observables
are measured. If only some of these observables are measured, valuable
information can still be extracted, with certain observables being particularly
sensitive to certain GPDs. In a second stage, we make a practical application 
of this code to the recent DVCS Jefferson Lab Hall A data from which we can 
extract numerical constraints for the two $H$ GPD Compton form factors.
\end{abstract}


\maketitle

\section{Motivation}

Generalized Parton Distributions (GPDs) have emerged during the past decade as a
powerful concept and tool to study nucleon structure. They describe,
among many other aspects, 
the (correlated) spatial and momentum distributions of the quarks in the nucleon 
(including polarisation information), its quark-antiquark content, a way
to access the orbital momentum of the quarks, etc. 

\begin{figure}[htb]
\epsfxsize=9.cm
\epsfysize=10.cm
\epsffile{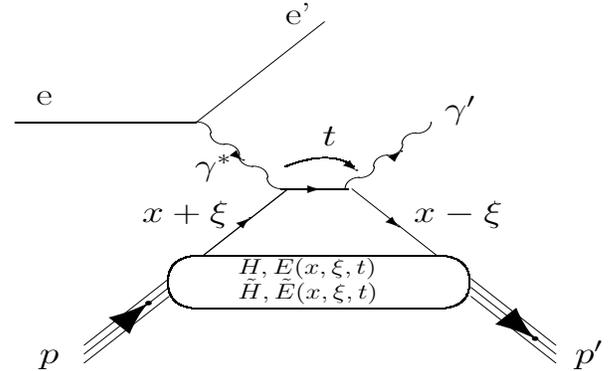}
\vspace{-3.2cm}
\caption{The handbag diagram for the DVCS process on the proton $ep\to e'p'\gamma '$. 
Here $x+\xi$ and $x-\xi$ are the longitudinal momentum fractions of the 
initial and final quark, respectively, and $t=(p-p')^2$ is the squared momentum 
transfer between the initial and final protons (or equivalently between the two
photons). There is also a crossed diagram which is not shown here.}
\label{fig:dvcs}
\end{figure}

Formally, in short, the GPDs are Fourier transforms of 
matrix elements in Quantum Chromo-Dynamics (QCD) for
light-cone bilocal operators between
nucleon states of different momenta. For helicity conserving quantities
in the quark sector, there are four GPDs, 
$H, \tilde H, E, \tilde E$ which depend, in leading order and leading twist QCD,
upon three variables: $x$, $\xi$ and $t$. Both $x$ and $\xi$ express the longitudinal
momentum fractions of the two quarks of the bilocal operator, 
while $t$ is the squared four-momentum transfer between the final and initial nucleon.
Experimentally, GPDs are most simply accessed through the measurement of
exclusive leptoproduction of a photon (Deep Virtual Compton Scattering -DVCS-) and, 
possibly, of a meson.
GPDs are then accessed through the factorization with an elementary perturbative
process, which leads to the so-called handbag diagram, predicted to be dominant at 
small $t$ and large $Q^2$, where $Q^2$ is the virtuality of the initial photon.
Fig.~\ref{fig:dvcs} shows schematically these notions for the DVCS process
on the proton on which, as a first step, we will concentrate our work and discussion
in this article. Indeed the leading twist contribution is expected to be the most 
directly accessible for DVCS and proton targets currently provide (and will also provide 
in the near future) the richest data set. We refer the reader to 
refs.~\cite{muller,ji,rady,collins,goeke,revdiehl,revrady} for the original theoretical 
articles and recent comprehensive reviews on GPDs for more details on the 
theoretical formalism of GPDs and DVCS.

Concerning actual data, the field is now rapidly growing. The first data
related to DVCS on the proton were the beam spin asymmetries (BSA) which have been published 
by the HERMES~\cite{dvcshermes} and CLAS~\cite{earlyclas} collaborations in 2001.
Although the kinematic range over which the data were integrated was large and, to some
extent, the actual exclusivity of the reaction could be disputed, these were the 
first very encouraging signals that the handbag diagram could be accessed,
since several theoretical predictions were in relatively good agreement with the data.
Then, longitudinally~\cite{shifeng} and transversely~\cite{hermest} polarized 
target spin asymmetries as well as beam charge 
asymmetries~\cite{elling} were measured which, though 
suffering from the same experimental limitations as the BSAs, 
also confirmed the general theoretical expectations. 

With these first encouraging signals found in non-dedicated experiments, a second 
generation of DVCS experiments has been launched in the past few years,
with fully dedicated detectors, i.e. high resolution electromagnetic calorimeters at 
Jefferson Lab (JLab) to detect the final-state photon and a recoil detector at HERMES
to detect the recoil proton. This has resulted in the publication of the first proton DVCS 
cross sections (beam-polarized and unpolarized) in the valence region at a few
precise kinematical points by the JLab Hall A 
collaboration~\cite{franck} and several BSAs over a large kinematic range
by the JLab Hall B collaboration~\cite{fx}. HERMES results with
the recoil detector are now eagerly awaited.

The field looks extremely promising with a series of novel experiments 
planned, in particular at JLab, aimed at accurately measuring new DVCS observables,
including: 
\begin{itemize}
\item Longitudinally polarized target spin asymmetries and cross sections (along 
with double polarizations observables) with JLab experiment E05114~\cite{E05114},
\item Transversely polarized target spin asymmetries and cross sections (along 
with double polarisations observables) with JLab experiment E08021~\cite{E08021},
\item New precise (unpolarized and beam-polarized) cross sections at new kinematics
with JLab experiments E06003~\cite{E06003} and E07007~\cite{E07007}. 
\end{itemize}
A similar experimental program in the longer term future is planned with the JLab 12 GeV 
upgrade, which will cover a larger phase space (see
JLab approved experiments E1206114~\cite{E1206114} and E1206119~\cite{E1206119}).

Anticipating this rich harvest of data, one asks the question how to extract 
the GPD information from them. We recall that the
DVCS process is accompanied by the Bethe-Heitler (BH) process, in which the 
final state photon is radiated by the incoming or scattered electron and not by the 
nucleon itself. The BH process, which is not sensitive to GPDs, is experimentally 
indistinguishable from the DVCS and interferes with it,
thus complicating the matter. Considering that the nucleon form factors are well-known 
at small $t$, the BH process is however precisely calculable theoretically. 

Another issue is that the GPDs, which are a function of the three variables 
$x$, $\xi$ and $t$, enter the DVCS amplitude within a convolution 
integral over $x$. Therefore, only $\xi$ and $t$ are accessible
experimentally (in the Bjorken limit, $\xi =\frac{x_B/2}{1-x_B/2}$ 
in which $x_B$ is the standard Bjorken variable). Formally, the DVCS amplitude is 
proportional to:
$\int_{-1}^{+1}d x {{H(\mp x,\xi,t)} \over {x \pm \xi \mp i \epsilon}}+...$
(where the ellipsis stands for similar terms for $E$, $\tilde{H}$ and $\tilde{E}$).
Decomposing this expression into a real and an imaginary part, we find that
the maximum information that can be extracted from the experimental 
data at a given ($\xi,t$) point is $H(\pm\xi,\xi,t)$, when measuring an observable
sensitive to the imaginary part of the DVCS amplitude, and $\int_{-1}^{+1}d x {H(\mp x,\xi,t) 
\over {x \pm \xi}}$, when measuring an observable sensitive to the real part of the DVCS
amplitude.

If ones reduces the range of $x$ from $\{-1,1\}$ to $\{0,1\}$ in the convolutions, 
there are in principle eight GPD-related quantities that can be extracted:
\begin{eqnarray}
&&P \int_0^1 d x \left[ H(x, \xi, t) - H(-x, \xi, t) \right] C^+(x, \xi),\label{eq:eighta} 
\\
&&P \int_0^1 d x \left[ E(x, \xi, t) - E(-x, \xi, t) \right] C^+(x, \xi),\label{eq:eightb} 
\\
&&P \int_0^1 d x \left[ \tilde H(x, \xi, t) + \tilde H(-x, \xi, t) \right] C^-(x,
\xi),\label{eq:eightc} 
\\
&&P \int_0^1 d x \left[ \tilde E(x, \xi, t) + \tilde E(-x, \xi, t) \right] C^-(x,
\xi),\label{eq:eightd} 
\\
&& H(\xi , \xi, t) - H(- \xi, \xi, t),\label{eq:eighte} \\
&& E(\xi , \xi, t) - E(- \xi, \xi, t),\label{eq:eightf} \\
&& \tilde H(\xi , \xi, t) - \tilde H(- \xi, \xi, t) \text{and}\label{eq:eightg} \\
&& \tilde E(\xi , \xi, t) - \tilde E(- \xi, \xi, t)\label{eq:eighth} 
\end{eqnarray}

with 
\begin{equation}
C^\pm(x, \xi) = \frac{1}{x - \xi} \pm \frac{1}{x + \xi}.
\end{equation}

\noindent
We will call these, respectively, in a symbolic notation, $Re(H)$, $Re(E)$,
$Re(\tilde{H})$, $Re(\tilde{E})$, $Im(H)$, $Im(E)$, $Im(\tilde{H})$
and $Im(\tilde{E})$. These are also often called the Compton Form 
Factors (CFFs). Note here the absence of $-\pi$ factors in our definition of
the $Im()$ CFFs with respect to ref.~\cite{kirch}.
In the following, for convenience and in a very loose way, we will speak of 
{\it real part} CFFs and {\it imaginary part} CFFs to designate generically the CFFs 
defined by eqs.~\ref{eq:eighta}-\ref{eq:eightd} and by eqs.~\ref{eq:eighte}-\ref{eq:eighth}, 
respectively and which correspond to the CFFs associated with the real 
and the imaginary parts of the DVCS amplitudes, respectively.
The CFFs can be decomposed into terms for individual quark flavors which, for the proton,
yields: $H(\xi , \xi, t)=\frac{4}{9}H^u(\xi , \xi, t)+
\frac{1}{9}H^d(\xi , \xi, t)$ and similarly for $E$, $\tilde H$ and $\tilde E$.
The quark flavor separation, which we will not tackle in this study, can be carried out
by measuring DVCS on a neutron target, which yields a different quark flavor
combination. The first measurements of the neutron BSA have been recently 
published~\cite{malek}.

In summary, given the interference of the BH process, the deconvolution issue regarding $x$ and
the large number (8) of independent quantities to be extracted from the data, it is clearly 
a non-trivial task to extract the GPDs from the experimental data and,
ultimately to map them in the three variables $x,\xi,t$. 

The first stage of any general program of measuring GPDs 
should certainly be to extract the eight CFFs from the data 
for a given $\xi,t$ point, in a model-independent way. 
This would be only the beginning of the program, since the $x$ dependence would still need 
to be deconvoluted using in principle a model with adjustable parameters. At this stage, 
let us mention that there might actually be a couple of ways around this issue: firstly, 
if the Double-DVCS 
process, (i.e. with a virtual photon in the final state) can be measured, then varying the
virtuality of the final state photon provides an extra lever arm and this allows us
to measure the GPDs at each $x,\xi,t$ values (though with some limitations if the final
photon is timelike)~\cite{ddvcs,ddvcs2}. Secondly, dispersion relations could
in principle reduce from eight to five the number of GPD quantities to be extracted, 
by expressing the real part CFFs defined by eqs.~\ref{eq:eighta}-\ref{eq:eightd} in terms 
of integrals over $\xi$ of the respective imaginary part CFFs defined 
by eqs.~\ref{eq:eighte}-\ref{eq:eighth}, plus a real subtraction constant (at fixed $\xi$ 
and $t$). This strategy requires one to 
measure data over a very wide range in $\xi$ (at fixed $t$) unless one has good reasons 
to truncate the integral or to extrapolate. We refer the reader to 
refs.~\cite{teryaev,ivanov,kumericki,marcmax} for discussions on this subject
and, in particular, to ref.~\cite{kumericki} for actual fits to the DVCS data in collider
kinematics. 

As a first approach, in this article, we focus our study on the most general case, 
in which the eight quantities of eq.~\ref{eq:eighta}-\ref{eq:eighth} are independent. 
Our present purpose is to understand to what extent, given the leading twist and 
leading order QCD DVCS amplitude, the CFFs can be extracted from 
various observables. We present our work in three steps, corresponding to the next three 
sections. In the next section, we present the general framework of the fitting code 
that we have developed. In particular, we present the tests that we have carried out
in a Monte-Carlo approach using simulated data. With ideal (i.e. insignificant) statistical 
error bars, we will learn
general features such as which observable is sensitive to which CFF, what are the highest 
reconstruction efficiencies that one can expect to achieve in such ideal conditions, etc. 
In the 
following section, we will simulate real experimental conditions by introducing a smearing of
the simulated data together with realistic error bars. We will then discuss the resulting reconstruction 
efficiencies and the uncertainties on the reconstructed GPD parameters.
Finally, in a last stage, in section~\ref{halla}, we will apply our fitting code to the 
recent JLab Hall A DVCS data and will attempt to extract quantitatively some of the
first real GPD quantities.

\section{Monte-Carlo study}
\label{acad}

In this section, we present the general features of our fitter code
and, in particular, the way we have tested it and established its reliability 
and efficiency. Our fitter code is simply based on the merging of the well
established VGG~\cite{vgg1,vgg2,gprv} code, which calculates numerically, using 
model CFFs, the leading order and leading twist handbag DVCS + BH amplitudes and 
observables, and the well known MINUIT minimization program from CERN~\cite{james}. 
The VGG code uses its own models for the GPDs but in fact any GPD model 
can be used. For our fitting purposes, the idea is to consider the
CFFs that enter the DVCS amplitude as free parameters to be fitted and to see how well 
they can be extracted from the DVCS Monte-Carlo (MC) data. 

We have started by testing and studying our code with pseudo-data. The procedure we 
have followed is based on a MC approach and can be summarized by the following 
general steps; for a given experimental kinematic point uniquely defined by 
$E_e$, $Q^2$, $\xi$ and $t$~($E_e$ being the beam energy):

\begin{enumerate}
\item We generate randomly a set of values for the quantities $Re(H)$, $Re(E)$, 
$Re(\tilde{H})$, $Re(\tilde{E})$, $Im(H)$, $Im(E)$, $Im(\tilde{H})$ and 
$Im(\tilde{E})$,
\item From these values, we calculate various observables 
which we will detail shortly as a function of $\phi$,
the azimuthal angle between the leptonic and hadronic planes, 
\item We fit with our code these distributions with eight ``parameters", which are 
meant to correspond to $Re(H)$, $Re(E)$, $Re(\tilde{H})$, $Re(\tilde{E})$, 
$Im(H)$, $Im(E)$, $Im(\tilde{H})$ and $Im(\tilde{E})$,
\item We finally compare the fitted values to the generated ones and draw
our conclusions.
\end{enumerate}

This procedure is going to be looped over several hundred times, exploring
the whole phase space of values, within given limits, that can be taken by $Re(H)$, $Re(E)$, 
$Re(\tilde{H})$, $Re(\tilde{E})$, $Im(H)$, $Im(E)$, $Im(\tilde{H})$ and 
$Im(\tilde{E})$. In doing so, we do not make any assumption or do not take into 
account any information 
on the absolute or the relative values of the generated GPDs, such as, for example, 
whether $Im(H)$ is dominant over $Im(\tilde{H})$ and $Im(\tilde{E})$, which could well be 
the case in reality. In this way, we place ourselves, in the framework
of this very general study, in the ``blindest", the least biased and the most conservative 
of situations. Entering physically motivated information or contraints on the CFFs
can of course only improve the results that we obtain.
Before presenting the results of our Monte-Carlo study, let us mention a few details.

We will consider nine independent observables
which can be expected to be measured with relatively good accuracy in the
near future. These are: $\sigma$, $\Delta\sigma_{z0}$, 
$\Delta\sigma_{0x}$, $\Delta\sigma_{0y}$, $\Delta\sigma_{0z}$, $\Delta\sigma_{zx}$, 
$\Delta\sigma_{zy}$, $\Delta\sigma_{zz}$ and $\Delta\sigma_{c}$. Here $\sigma$ refers 
to the unpolarized cross section. When there are two indices, $\Delta\sigma$ 
refers to the difference of polarized cross sections
and the two indices refer respectively to the polarization of the 
beam and of the target (i.e. $\Delta\sigma_{z0}$ is the difference of
cross sections with a longitudinally polarized beam and an unpolarized target, as 
has been measured recently by the JLab Hall A collaboration~\cite{franck},
and $\Delta\sigma_{zz}$ corresponds to
the difference of cross sections that can be obtained with a longitudinally polarized beam 
and a longitudinally polarized target). Finally, $\Delta\sigma_{c}$ refers to the difference of
unpolarized cross sections between a negative and a positive lepton beam, i.e.
proportional to the beam charge asymmetry measured by HERMES~\cite{elling}.

Regarding the parameters to be fitted,
as a first approach, we have taken into account only seven CFFs, instead of eight: 
$Re(H)$, $Re(E)$, $Re(\tilde{H})$, $Re(\tilde{E})$, $Im(H)$, $Im(E)$, $Im(\tilde{H})$, 
i.e. we have set $Im(\tilde{E})$ to $0$. The reason is that the GPD $\tilde{E}$ is usually 
associated with the pion pole $t$-channel exchange which is real.
Nothing keeps us in principle from considering $Im(\tilde{E})$ and taking
it as an extra free parameter in the fit. However, there
is also clearly no need to complicate the fit procedure if it
is not justified. Conversely, we have decided to let 
$Re(\tilde{E})$ be a free parameter, even though it is supposed to 
reflect the pion pole and is therefore well-constrained. In this way, the conjecture that 
$\tilde{E}$ comes from the pion pole can be verified. If this conjecture proves false, 
the whole parametrization of $\tilde{E}$
will clearly have to be revisited including its imaginary
contribution. Thus $Im(\tilde{E})=0$ is the only model assumption that we will make 
in this study.

Also, we have decided to take as fit parameters, not the CFFs of 
eq.~\ref{eq:eighta}-\ref{eq:eighth} themselves, but the deviations with respect to them.
This is not an assumption, simply a convention.
In other words, we take some (supposedly) realistic reference values 
for $Re(H)$, $Re(E)$, $Re(\tilde{H})$, $Re(\tilde{E})$, $Im(H)$, $Im(E)$ 
and $Im(\tilde{H})$ and fit the coefficients that multiply these reference 
values to the MC data. Therefore in our MC study, we are going to generate 
randomly, in a given range, seven real numbers that multiply the seven reference CFFs.
From these new (random) CFFs, we generate the DVCS observables. We fit them 
and our aim is to recover the seven initial multiplicative coefficients,
knowing the numerical values of the reference CFFs. We will call these multiplicative 
coefficients, the GPD multipliers and denote them as $a(Re(H))$, $a(Re(E))$, 
$a(Re(\tilde{H}))$, $a(Re(\tilde{E}))$, $a(Im(H))$, $a(Im(E))$ and $a(Im(\tilde{H}))$.
Similarly to the CFFs, in the following, for convenience, we will speak
of {\it real part} and {\it imaginary part} multipliers, although, mathematically,
all these numbers are obviously real.
These can be interpreted as ``ratios" of the fitted CFFs to the reference CFFs.
Clearly, this is only a matter of convention and instead of these GPD multipliers, 
we could certainly have generated and fitted $Re(H)$, $Re(E)$, etc. directly. Our general 
motivation is to create the most efficient and robust code by 
starting the minimization procedure as close as possible to the true
solution. This implies that the reference CFFs must be as realistic as possible.
For these reference CFFs, we have taken, as a first approach, those given by the VGG 
parametrization. It is by no means infered that the VGG parametrization is
the most realistic one that currently can be found in the literature,
but this was obviously the
most convenient choice in our case because they provide a decent 
description of several existing DVCS experimental data~\cite{fx,dvcshermes,earlyclas}. 
However, it is clear 
that instead of VGG as a reference, any other GPD model could be taken.



In the present study, we have generated seven numbers $a(Re(H))$, $a(Re(E))$, 
$a(Re(\tilde{H}))$, $a(Re(\tilde{E}))$, $a(Im(H))$, $a(Im(E))$ and $a(Im(\tilde{H}))$.
The range we have considered is $\{-4,4\}$. This means that we allow the randomly generated 
CFFs to vary in absolute value up to a factor $4$ from the VGG value. 
This range $\{-4,4\}$ is quite arbitrary and does not have any
significant impact on the subsequent results of this section. Whether such large
variations from the VGG CFFs are realistic or not is not particularly the object of this 
section which is meant as an exercice, testing the ability to extract GPDs. Let us just 
mention that the GPDs of the VGG code are normalized and respect the 
various model-independent relations ($H^q(x,0,0)=q(x)$, 
$\tilde{H}^q(x,0,0)= \Delta q(x)$, where $q(x)$ and $\Delta q(x)$ are
the unpolarized and polarized parton distributions functions of flavor $q$, 
respectively) and sum rules 
($\int_{-1}^{+1} d x H^{q}(x,\xi,t) \,=\, F_1^{q}(t)$, etc.). Although
these relations are quite constraining and do not permit a fully arbitrary
normalisation of the GPDs, let us note that there are contributions to the GPDs
which can escape any formal normalization constraints. For instance, the 
so-called $D$-term~\cite{weiss}, which cancels at $\xi$=0 and is odd in $x$
so that it does not contribute to the form factor sum rule, is such
a contribution. Also, GPD parametrizations based
on other approaches such as in ref.~\cite{diquark} can find in the valence region
an $H$ GPD larger by a factor 2 to 3 compared to VGG.


Once the seven initial multipliers are generated, the next step
is to calculate with the VGG code the DVCS+BH amplitude
and the various (unpolarized, singly polarized, doubly polarized)
resulting cross sections. We recall that, in this study, we restrict ourself to 
leading-twist. However, in order to ensure gauge invariance of the DVCS amplitude, 
we mention that the VGG amplitude contains a small (on the order of 10\%
depending on the kinematics) twist-3 contribution~\cite{vgg1,vgg2}. Since
Wandura-Wilczek kinematical twist-3~\cite{kivel} and some other contributions
to higher twists are also available in the VGG code ($k_\perp$ effects
for instance)~\cite{vgg1,vgg2}, let us mention that fits can also be done
beyond leading twist at the cost of longer computing time.

\begin{figure}[htb]
\epsfxsize=9.cm
\epsfysize=10.cm
\epsffile{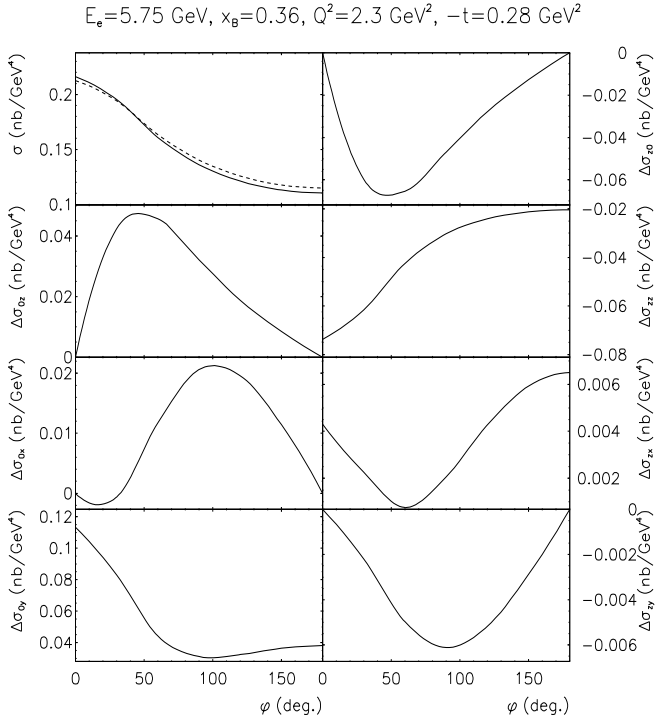}
\caption{The nine independent DVCS observables considered in this work as a function
of $\phi$, at the kinematics $E_e$=5.75 GeV, $x_B$=0.36, $Q^2$=2.3 GeV$^2$ and 
$-t$=0.28 GeV$^2$. The curves have been generated with the GPD multipliers of 
table~\ref{tab:param}.
The left column, from top to bottom, corresponds to the unpolarized cross section 
with an electron beam (solid curve) and a positron beam (dashed curve) and the three target single polarization
differences of cross sections: $\Delta\sigma_{0z}$, $\Delta\sigma_{0x}$ and
$\Delta\sigma_{0y}$.
The right column, from top to bottom, corresponds to the beam polarized difference of cross section 
$\Delta\sigma_{z0}$ and the three beam-target double polarization differences of cross 
sections: $\Delta\sigma_{zz}$, $\Delta\sigma_{zx}$ and $\Delta\sigma_{zy}$.
All observables are four-differential in $dx_BdQ^2dtd\phi$.}
\label{fig:example}
\end{figure}

At given $E_e$, $\xi$, $Q^2$ and $t$ values, we generate the $\phi$
distributions of the nine observables previously mentioned. 
Fig.~\ref{fig:example} shows such distributions calculated from
the set of randomly generated $a()$ multipliers which are displayed in 
table~\ref{tab:param}. The kinematics correspond to that of the recent 
JLab Hall A data~\cite{franck}: $E_e$=5.75 GeV, $x_B$=0.36 (corresponding to $\xi$= 0.22), 
$Q^2$=2.3 GeV$^2$ and $-t$=0.28 GeV$^2$.
Our goal is therefore to recover the $a()$ values of table~\ref{tab:param} 
from the fit to the $\phi$ distributions of fig.~\ref{fig:example}.
Once again, we recall that these values are the coefficients that multiply the 
VGG CFFs, i.e. not the CFFs themselves. 
The corresponding VGG CFFs are also displayed in table~\ref{tab:param}.
The VGG code has several options for the parametrization of the GPDs.
As we said, it does not really matter which particular option is used since
the VGG CFFs are just meant to be used as starting or reference values and
what is ultimately extracted are the deviations from these values.
We have used the Regge inspired unfactorized
ansatz for the $t$-dependence~\cite{gprv} of $H$ and $E$, a factorized
$t$ dependence ansatz for $\tilde{H}$ and the pion pole for the modelling
of $\tilde{E}$. Also, no $D$-term was used and the
parameters $b_{val}=b_{sea}=1$ in the profile function of the Double Distributions
were selected.
 
Obviously, since the GPD multipliers of 
table~\ref{tab:param} have been randomly generated for our 
study, they have no particular meaning, and 
the curves of fig.~\ref{fig:example} have no relation to the actual 
$\sigma$ and $\Delta\sigma_{z0}$ that the JLab Hall A collaboration has
measured at similar kinematics. 
 
\begin{table*}[htb]	
\begin{center}	
\begin{tabular}{|l|c|c|c|c|c|c|c|}
\hline	
&$Re(H)$  & $Re(E)$ & $Re(\tilde{H})$ & $Re(\tilde{E})$ & $Im(H)$ & $Im(E)$ & $Im(\tilde{H})$\\
\hline \hline
$a()$  & 0.378 & -1.818 & 3.296 & -0.699 & -3.732 & 2.608 & -1.285 \\
VGG value & 0.658 & 0.355 & 0.458 & 41.705 & 1.58 & 0.48 & 0.43 \\
\hline
\end{tabular}	
\caption{One set of seven randomly generated GPD multipliers $a()$ which are used in the 
illustration of our study, together with the corresponding VGG values for the seven CFFs
at $x_B$=0.36 (corresponding to $\xi$= 0.22), and $-t$=0.28 GeV$^2$. } 
\label{tab:param}
\end{center}	
\end{table*}	

In this section, we generate the $\phi$ distributions with ``ideal" error bars, 
i.e. 5$\times$10$^{-4}$ in relative value to the cross sections. The idea
is first to gain confidence in our fitting program as well
as to understand, under ideal conditions, some general features
of the fit procedure such as which GPD is sensitive to which observable,
the number of observables needed to extract all GPDs, and so forth.
In the following section, we will discuss real conditions by 
assigning realistic error bars to these distributions. 

In fitting these $\phi$ distributions with the
seven free parameters $a(Re(H))$, $a(Re(E))$, $a(Re(\tilde{H}))$, $a(Re(\tilde{E}))$, 
$a(Im(H))$, $a(Im(E))$ and $a(Im(\tilde{H}))$, we stress that
we ignore any information that we know about the initial generated values,
the idea being to match as much as possible real conditions. In this study,
we set the starting values to 0. For real data,
as we said, in order to be as close as possible to the true solution,
the starting values should be set to 1.
The range over which the values of the seven coefficents are allowed to vary
is set to $\{-5,5\}$, i.e. somewhat larger than the range
of the generated values. This point 
is essentially the only piece of information that we take from the initial 
input.
We have used a least square method and the MIGRAD minimizer of MINUIT to 
perform the fit. The quantity that we minimize is thus:

\begin{equation}
\chi^2=\sum_{i=1}^{n}
\frac{(\sigma^{theo}_i-\sigma^{exp}_i)^2}{(\delta\sigma^{exp}_i)^2}
\label{eq:chi2}
\end{equation}

\noindent where $\sigma_{theo}$ is the theoretical cross section (or difference of 
cross sections)
from the VGG code, $\sigma_{exp}$ is the corresponding (simulated) experimental 
value (generated from the VGG code as well) and $\delta\sigma_{exp}$ 
is its associated experimental 
error bar. We recall that in this section, we consider ideal error bars.
The index $i$ runs over the number of observables
to be fitted, i.e. nine at maximum in the present context.
For all the results that we present in the next two sections of this article, 
fits have been carried out
with seven experimental points in $\phi$ spread in steps of $\phi$=30$^\circ$ between
0$^\circ$ and 180$^\circ$. Little improvement is observed in fitting more $\phi$ 
values however, computing time increases. Fitting seven experimental points for each of 
the nine observables 
with our seven parameter formulation takes about thrity minutes, running on 
an average computer. Fitting only $\sigma$ and $\Delta\sigma_{z0}$ takes 
less than ten minutes.

\begin{figure}[htb]
\epsfxsize=9.cm
\epsfysize=10.cm
\epsffile{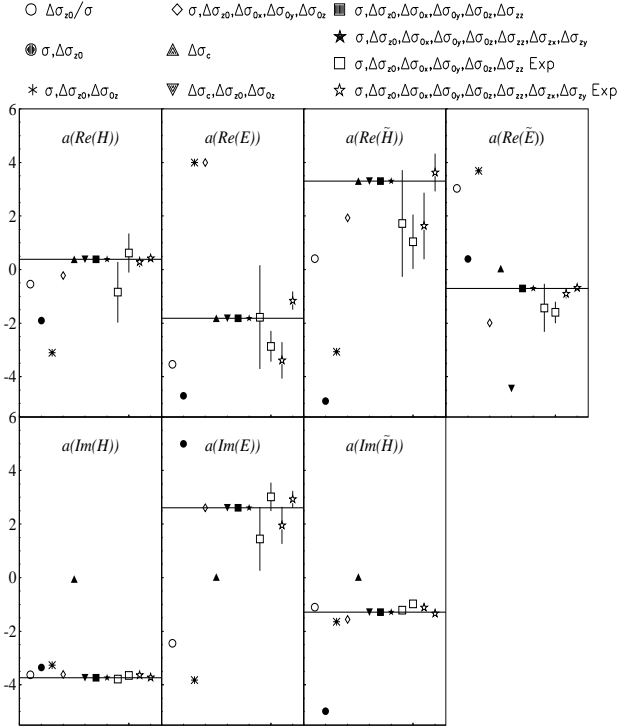}
\caption{One particular example of the fit results for different topologies. 
The solid line indicate the generated values of the a() GPD multipliers of 
table~\ref{tab:param}.}
\label{fig:events}
\end{figure}

\begin{figure}[htb]
\epsfxsize=9.cm
\epsfysize=10.cm
\epsffile{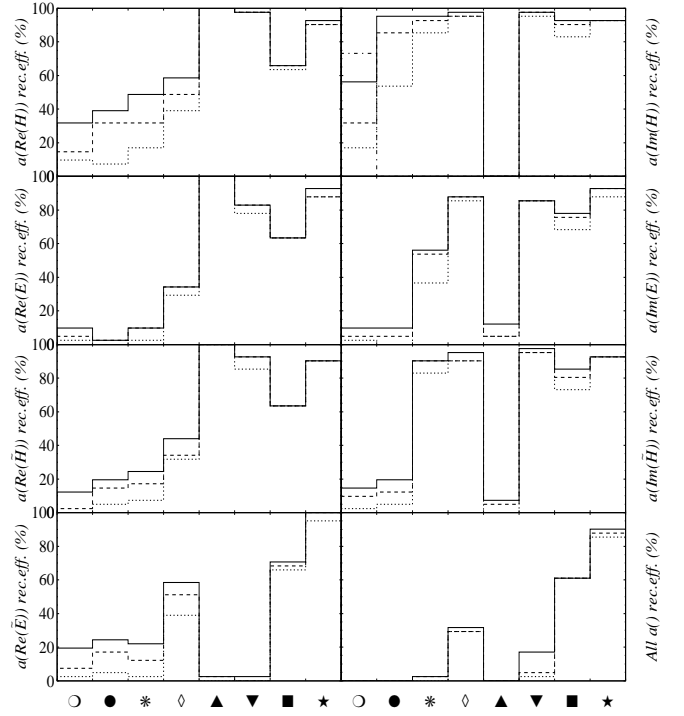}
\caption{Reconstruction efficiencies for all the topologies
of the seven GPD multipliers and of all of seven in concert. 
The symbols coresponding to the topologies have been defined in fig.~\ref{fig:events}.
Right bottom plot: reconstruction
of all GPD multipliers simultaneously. The reconstruction efficiency is defined
as the ratio of the number of well reconstructed events to the number of generated 
ones. An event is ``well reconstructed" when the reconstructed numerical value
matches the generated one to within 5\% (dotted curve), 15\% (dashed curve)
or 25\% (solid curve). The dash-dotted curve for the $Im(H)$ BSA corresponds 
to the reconstruction efficiency (at the 15\% level) if the starting value 
for $Im(H)$ in MINUIT is set to the generated one $\pm 20$\%.}
\label{fig:eff}
\end{figure}

We have generated
several hundred events in order to have statistically significant results
where an event is understood to be a set of seven real values 
$a(Re(H))$, $a(Re(E))$, $a(Re(\tilde{H}))$, $a(Re(\tilde{E}))$, 
$a(Im(H))$, $a(Im(E))$ and $a(Im(\tilde{H}))$. Let us now see how the
fitted values of the GPD multipliers compare to the generated ones.
We present our results with the support of two figures. 
In the following, we call a topology a particular combination of some of the nine observables:
$\sigma$, $\Delta\sigma_{z0}$, $\Delta\sigma_{0x}$, $\Delta\sigma_{0y}$, 
$\Delta\sigma_{0z}$, $\Delta\sigma_{zx}$, $\Delta\sigma_{zy}$, 
$\Delta\sigma_{zz}$ and $\Delta\sigma_{c}$.
Fig.~\ref{fig:events} compares the generated and the reconstructed
values of our seven GPD multipliers under different topologies, 
using one particular event corresponding
to the generated GPD multipliers of table~\ref{tab:param}. 
Fig.~\ref{fig:eff} summarizes our results for several hundred events,
in the phase space $\{-4,4\}$ of the $a()$ multipliers,
for each of the fitted parameters, under the different topologies of 
fig.~\ref{fig:events}.
The particular event of fig.~\ref{fig:events} has been chosen for illustrative
purposes because it allows us to make several (not all, though) of the conclusions that we 
can draw with hundreds of events. Let us start by discussing this figure.
 
In fig.~\ref{fig:events}, the solid lines show the (randomly) generated value
of the seven multipliers (see table~\ref{tab:param}), while all the points show 
the fitted (reconstucted) values obtained from our fitting program, under different topologies.
We are going to discuss them one by one, in parallel with figure~\ref{fig:eff},
to understand if the features that we can infer from fig.~\ref{fig:events} are 
general or simply an accident. 

In fig.~\ref{fig:events}, the empty circles show the reconstructed values
obtained if only one observable, the DVCS BSA $\frac{\Delta\sigma_{z0}}{\sigma}$,
(which is so far the most widely measured observable) is fitted. 
We see in this figure that only $a(Im(H))$ is well 
reconstructed, as well as $a(Im(\tilde{H}))$, to a somewhat lesser extent.
On very general grounds, it is not at all surprising that fitting only 
one $\phi$ distribution, i.e. the BSA here,
with seven parameters, we are not able to recover all of them. 
From fig.~\ref{fig:eff}, we see that the good reconstruction
of $a(Im(H))$ is a relatively general feature while the reconstruction of 
$a(Im(\tilde{H}))$ is more of a statistical accident.
Fig.~\ref{fig:eff} shows, for several hundred randomly generated events, the
ratio of the number of well-reconstructed events to the number of generated 
ones, which we call a reconstruction efficiency, for each GPD multiplier. 
``Well reconstructed" means that the reconstructed numerical value 
of a multiplier falls within 5\% (dotted curve), 15\% (dashed curve)
or 25\% (solid curve) of the generated value. One learns from
this figure that by fitting only the BSA one recovers, at the 25\% level, 
about 55\% of the generated values of $a(Im(H))$ while only 
$\approx$ 30\% and less than 20\% if the reconstructed values
are within 15\% and 5\% of the generated ones, respectively (first column
of the upper right plot of fig.~\ref{fig:eff}). Once again, the efficiencies 
for recovering $a(Im(H))$ are calculated with all $a()$'s in the 
range $\{-4,4\}$. Therefore, if $a(Im(H))$
happens to be close to zero while the other $a()$'s not, it shouldn't be 
surprising that $a(Im(H))$ is not well-recovered. If in reality $Im(H)$ dominates 
over the other GPDs then these efficiencies can only increase. Therefore, the numbers
that we quote for these efficiencies and which are displayed in
fig.~\ref{fig:eff} should be taken with great care because they really depend 
on the relative value of each CFF. In addition, the kinematics 
can also change the weight of each CFF. Thus these efficiencies 
reflect more a ``sensitivity" of a given CFF to some observable(s) rather 
than an absolute efficiency and they have only a general statistical meaning 
in the present context.

Another way to improve the efficiency of the fit is to set 
the starting values close to the true solution. It will then be less likely
that MINUIT falls into some local minimum.
For instance, the dash-dotted curve in the first column (i.e the BSA)
of the upper right plot (i.e $Im(H)$) of fig.~\ref{fig:eff} shows the 
reconstruction efficiency (at the 15\% level) if the starting value in
MINUIT for $Im(H)$ is set to the \underline{generated} $Im(H)$ (smeared at the
20\% level) instead of at 0 as we have done so far. 
Biasing and guiding the fit in such a way more than
doubles the probability of recovering the original value.

Despite all these caveats, we can still learn a lot from fig.~\ref{fig:eff}.
We can note that 
$a(Im(\tilde{H}))$ is not recovered more than 15\% of the time,
which means that the good reconstruction noted
in fig.~\ref{fig:events} is not generally the case and that
the reconstruction of the $Im(\tilde{H})$ CFF in this topology is 
thus very dependent on its particular value and weight with respect to the other
CFFs. On the contrary, fig.~\ref{fig:eff} shows a feature that fig.~\ref{fig:events} might
not have hinted to, i.e. that the BSA has some non-negligible
sensitivity to $Re(H)$ since in $\approx$ 35\% of the cases, it
is well-reconstructed at the 25\% level. 

The sensitivity of the BSA 
to $Im(H)$ is a well-known feature since it can be shown analytically
(see ref.~\cite{kirch} for instance) that $Im(H)$ is a dominant contribution 
to the difference of beam polarized cross sections $\Delta\sigma_{z0}$ which is 
the numerator of the BSA. The (limited) sensitivity of the BSA to 
$Re(H)$ can be understood through its contribution to the denominator
of the BSA which is the unpolarized cross section ${\sigma}$ 
where the real part CFFs enter.

In fig.~\ref{fig:events}, the solid circles show the reconstructed 
values which are obtained for our seven parameters, if now the two observables just discussed, 
the unpolarized cross section $\sigma$ and the difference of beam polarized cross section 
$\Delta\sigma_{z0}$, are both used in the fit rather than just their ratio. 
In fig.~\ref{fig:events}, we see
similar features to the previous case where only the BSA was fitted, i.e. that
only $a(Im(H))$ is fairly well recovered: -3.354 reconstructed against
-3.732 generated (table~\ref{tab:param}). It also confirms that the good
reconstruction of $a(Im(\tilde{H}))$ in the case of the BSA 
was more an accident than a solid feature. However, the advantage 
of fitting simultaneously these two observables instead of simply 
their ratio is shown well in fig.~\ref{fig:eff} where we see that the reconstruction
efficiency of $a(Im(H))$ now reaches more than 85\% (at the 15\% precision level)
while it was less than 35\% in the BSA topology. This illustrates the
strong correlation between $\Delta\sigma_{z0}$ and $Im(H)$ and that $Im(H)$ 
can be recovered, through this topology essentially
independently of the values of the other GPDs. This conclusion that 
$\Delta\sigma_{z0}$ is mostly 
sensitive to $Im(H)$ and barely to the other GPDs is not new and was
already pointed out for instance in ref.~\cite{kirch} analytically.
We confirm this numerically.

The topology marked with an asterisk in fig.~\ref{fig:events} is the result 
of the fit when adding one more observable. In addition to the unpolarized cross
section $\sigma$ and the difference of beam polarized cross section $\Delta\sigma_{z0}$, we
now also fit the difference of longitudinally polarized target
cross sections $\Delta\sigma_{0z}$. One sees in fig.~\ref{fig:events}
that the four real part CFFs 
are as poorly reconstructed as in the previous cases. However,
now there is a second imaginary part CFF, besides $a(Im(H))$, that is 
well-recovered: $a(Im(\tilde{H}))$=-1.009 in the fit compared to
the generated value -1.285 (table~\ref{tab:param}). We see from fig.~\ref{fig:eff} that
this conclusion, i.e. the strong sensitivity of $\Delta\sigma_{0z}$ to 
$Im(\tilde{H})$, which was also already pointed out in ref.~\cite{kirch}, 
is very general. In more than 80\% of the cases, $a(Im(\tilde{H}))$
is well reconstructed in this topology. We also note that the
precision on $a(Im(H))$ is significantly improved.
The reconstruction efficiencies corresponding to the 5\% (dotted
curve) and 15\% (dashed curve) criteria have increased (up to
$\approx$ 80\% compared to $\approx$ 50\% in the previous topology
for the dashed curve).

We now bring two more observables into the fit, i.e. the difference 
of cross sections obtained with the transverse $x$ and $y$ transverse polarizations 
of the target. The corresponding results are shown by
the open diamonds in fig.~\ref{fig:events}. We observe that the three 
imaginary part CFFs are now well-reconstructed and that this conclusion
remains true generally. Fig.~\ref{fig:eff} shows that 
in more than 80\% of the cases, $a(Im(H))$, $a(Im(\tilde{H}))$ 
and $a(Im(E))$ are well-reconstructed in this topology. 
We can thus infer that the target transverse (single) polarisation observables 
are very sensitive to $Im(E)$ and that measuring the five observables
$\sigma$, $\Delta\sigma_{z0}$, $\Delta\sigma_{0x}$, $\Delta\sigma_{0y}$, 
and $\Delta\sigma_{0z}$ allows us to extract the three imaginary part CFFs reliably.
Furthermore, fig.~\ref{fig:eff} shows that, although it is not 
illustrated particularly well by fig.~\ref{fig:events}, the combination of 
these five observables has a significant sensitivity to the four real part 
multipliers. About 40\% of them are reconstructed well.

Next, the upright solid triangles show the results of fitting \underline{only}
the beam charge difference of cross sections (i.e. no polarization observables). 
We see now that
the three real part multipliers $a(Re(H))$, $a(Re(E))$, $a(Re(\tilde{H}))$ are 
reconstructed well while no imaginary part multiplier is recovered at all.
Fig.~\ref{fig:eff} confirms that this is the case at basically the 100\% level
(the few well-reconstructed imaginary part multipliers in this topology
in fig.~\ref{fig:eff} are clearly accidental).
Because, on general grounds, $\Delta\sigma_{c}$ is expected to be sensitive only to 
the real part 
of the DVCS amplitude, this result comes as no surprise. We also note 
in fig.~\ref{fig:events} that the real part multiplier
$a(Re(\tilde{E}))$ is not reconstructed in this topology. This is a general feature 
since, as seen from fig.~\ref{fig:eff}, $a(Re(\tilde{E}))$ is not
reconstructed in any of the cases with our high statistics 
event sample. Again, the couple of percent of events well-reconstructed clearly 
result by coincidence. This feature can be demonstrated analytically by 
the same kind of argument used to show that the difference of beam
polarized cross sections $\Delta\sigma_{z0}$ is insensitive to
$Im(\tilde{E})$.

In fig.~\ref{fig:events}, the upside-down solid triangles show the effect of 
fitting $\Delta\sigma_{z0}$ 
and $\Delta\sigma_{0z}$, in addition to $\Delta\sigma_{c}$. 
Now, in addition to the three real part multipliers $a(Re(H))$, $a(Re(E))$, $a(Re(\tilde{H}))$, 
the three imaginary part multipliers are well reconstructed following our previous discussions. 
We note that there is a slight decrease of the reconstruction efficiency
of the three real part multipliers when fitting simultaneously $\Delta\sigma_{c}$,
$\Delta\sigma_{z0}$ and $\Delta\sigma_{0z}$ compared to fitting only
$\Delta\sigma_{c}$. This is due to the fact that, here, we have minimized
the sum of the three $\chi^2$'s corresponding to the minimization of $\Delta\sigma_{c}$,
$\Delta\sigma_{z0}$ and $\Delta\sigma_{0z}$, and it can be shown that the derivatives
with respect to the imaginary part CFFs are the highest compared to
those with respect to the real part CFFs. In other words, in doing so,
one loses some sensitivity to the real part CFFs. Therefore, in such a situation,
where a set of observables is sensitive only to some other independent CFFs, 
like $\Delta\sigma_{c}$
to the real part CFFs, and another set of observables only to some other independent 
CFFs, like $\Delta\sigma_{z0}$ and $\Delta\sigma_{0z}$ to the imaginary part CFFs,
instead of fitting simultaneously both sets, one should obviously adopt a two-step procedure 
and fit independently each set of observables. To illustrate this trivial effect,
we leave in fig.~\ref{fig:eff} the results of our simultaneous fit, with
the understanding that in this topology, if we had fitted separately $\Delta\sigma_{c}$, we would 
have recovered the same efficiencies of the upright solid triangles topology for the real part CFFs.
This being said, we then see that six of the seven GPD multipliers can be reconstructed
in a very economical way by measuring only three observables, and in particular,
that the gold-plated way to access the real part CFFs is with the beam charge difference of cross 
sections.

However, the next symbol (solid square) in fig.~\ref{fig:events} shows that there is 
another way of accessing the real part multipliers with high efficiency. The solid-square 
topology contains the double-polarization 
observable $\Delta\sigma_{zz}$ in addition to the open-diamond
topology, which was made up of the unpolarized cross section and the 
four single beam or target polarization
observables. We then see that all real part multipliers (in particular including 
$a(Re(\tilde{E}))$, to which $\Delta\sigma_{c}$ was insensitive) and 
imaginary part multipliers are well recovered. Fig.~\ref{fig:eff} confirms
that the particular event of fig.~\ref{fig:events} is not an accident
and that these conclusions can be generalized. In about 80\% of the
cases, the imaginary part multipliers are well-reconstructed
at the 25\% level, similarly for higher precisions,
and in more than 65\% of the cases for the real part multipliers.

The solid stars show 
the results obtained when fitting all polarization observables 
simultaneously, i.e. now adding the two remaining double 
polarization observables $\Delta\sigma_{zx}$ and $\Delta\sigma_{zy}$
to the previous topology. We see 
from fig.~\ref{fig:events} that all seven parameters are well-reconstructed, 
as in the previous topology, but fig.~\ref{fig:eff} shows that this
is now at a higher efficiency which reaches $\approx$
90\% for all seven multipliers. Fitting eight observables with seven parameters
overconstraints the fit, with redundant information,
and therefore significantly improves the efficiency.

The remaining symbols in fig.~\ref{fig:events} will be discussed in
the next section. So far, we have described the general features 
of our fitting program in the ideal situation,
i.e. with perfectly precise data, and have
demonstrated its general reliability and power to recover all of our 
seven arbitrary randomly generated parameters (CFFs) from fits to 
the $\phi$ distributions, 
given enough observables. We learned which observables are sensitive
to particular CFFs and that, if only some of these observables
are available, we could still extract some specific CFF.
In particular, one can access in principle the real part CFFs
with relatively high efficiency using double polarisation observables
in addition to beam charge differences.

We now turn to realistic conditions. Ultimately, this program has to 
be used on real data with finite precision.
In the next section, we therefore simulate real experimental conditions 
and find how robust and reliable the program remains.

\section{A more realistic study}

We are now going to simulate realistic pseudo-data by assigning experimental
error bars to the data to be fit. 
The procedure consists not only in assigning an error bar to the simulated 
$\phi$ data points but also in smearing the 
central value according to a gaussian probability distribution whose standard deviation 
is equal to the error bar.  

Inspired from refs.~\cite{franck,E06003}, we will assign, at all
$\phi$'s, a 3.5\% error bar to the unpolarized cross section. For all
the differences of cross sections, we will consider two cases: 10\% or
5\%, the former being approximatively the error bar on $\Delta\sigma_{z0}$
which is quoted in refs.~\cite{franck,E06003} and the latter being a slightly 
more challenging experimental goal.

We return to fig.~\ref{fig:events}. The open square and open star symbols show the result of
the two last topologies discussed in section~\ref{acad} (i.e. the solid
square and solid stars) when realistic error bars are applied
to the pseudo-data in the $\phi$ distributions. In each of the seven plots, there are two
open squares and two open stars.
The leftmost, for both symbols, is the result of the fits when the differences
of cross sections are smeared by 10\% and the rightmost when they are smeared by 5\%. This can clearly
be inferred from the size of the error bar of the fitted parameters which is systematically
smaller when the data are smeared by 5\% compared to 10\%. The uncertainties on the fitted parameters
that are presented here are the quadratic errors from MINUIT, which is 
sufficient in the present context where we are looking at statistical effects. 
However, we will see in the
next section when fitting the JLab Hall A $\sigma$ and $\Delta\sigma_{z0}$ real data
that the error determination requires a more dedicated study, especially when dealing
with a underdetermined problem with more parameters to fit than available observables.

We see in fig.~\ref{fig:events}, for this particular event, that the general conclusions that we 
reached in the previous section remain, i.e. that essentially all seven GPD
multipliers are relatively well-recovered when fitting the six observables $\sigma$, $\Delta\sigma_{z0}$, 
$\Delta\sigma_{0x}$, $\Delta\sigma_{0y}$, $\Delta\sigma_{0z}$ and $\Delta\sigma_{zz}$
(open square topology) and eight observables with $\Delta\sigma_{zx}$ and $\Delta\sigma_{zy}$ in addition 
(open star topology). However, while in the previous section,
the original central values were recovered almost perfectly in most of the cases, there
is now a clear dispersion of the reconstructed central values with respect
to the generated central ones. We note that this difference
is always within two standard deviations and therefore that the errors
estimation provided by MINUIT seems very reasonable. 

\begin{figure}[htb]
\epsfxsize=9.cm
\epsfysize=10.cm
\epsffile{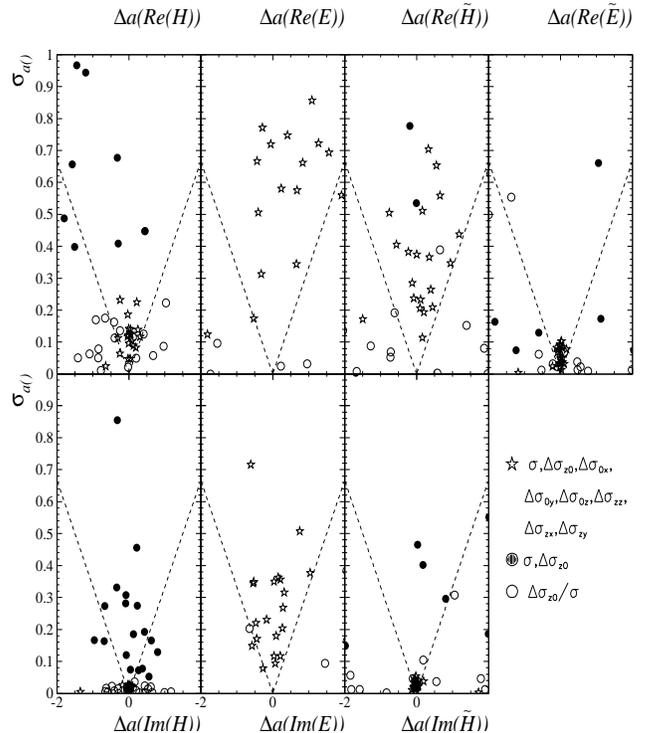}
\caption{Uncertainty on the fitted parameters $\sigma_{a()}$, as given by MINUIT, 
as a function of the difference between the fitted and generated values, for
the open-star (with 10\% error bar for the pseudo-data), solid-circle and open-circle topologies. 
The dashed line indicates the three $\sigma_{(a)}$ limits.}
\label{fig:sig}
\end{figure}

Fig.~\ref{fig:sig} confirms this conclusion with a larger sample (though limited in the 
display in order not to overcrowd the figure) of events where we observe that 
for basically all open stars events (with 10\% error bar for the pseudo-data) the difference 
between the generated and 
reconstructed values $\Delta a()$ is within three $\sigma_{a()}$'s, with $\sigma_{a()}$ 
being the MINUIT
uncertainty previously discussed. The few cases where it is not the case
can easily be identified after inspection; either the $\chi^2$ of the corresponding fit
is very bad or the generated value is close to 0,
meaning that the observables had very little sensitivity to it.

Fig.~\ref{fig:sig} also shows the solid circles resulting from fits
of only $\sigma$ and $\Delta\sigma_{z0}$. We note that these symbols are visible essentially
only in the $Re(H)$ and $Im(H)$ plots. This confirms the observations that we made
in the previous section that this topology is
sensitive to only these two GPD quantities (predominantly $Im(H)$). Indeed, the absence
of solid circles on the other plots means that their uncertainty $\sigma_{a()}$ is 
off-scale, indicating the total lack of sensitivity of this topology
to these GPD quantities. For this topology, we also observe the larger $\sigma_{a()}$ 
values in average for
$Re(H)$ than for $Im(H)$ giving evidence for the stronger sensitivity of the solid
circle topology to $Im(H)$. 

\begin{figure}[htb]
\epsfxsize=9.cm
\epsfysize=10.cm
\epsffile{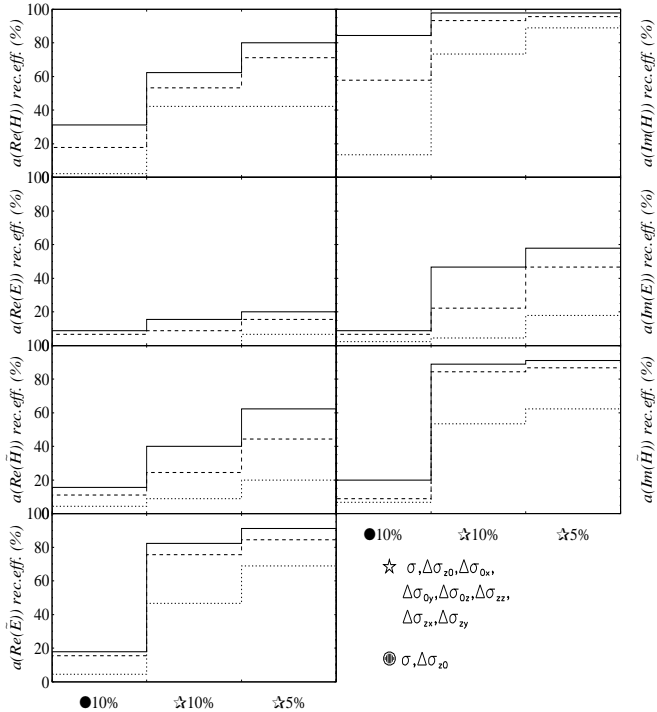}
\caption{Reconstruction efficiencies of three ``experimental" topologies
for the seven GPD multipliers. Left: efficiency of the fit of $\sigma$ 
and $\Delta\sigma_{z0}$ with the experimental $\phi$ distributions being smeared
by 3.5\% for $\sigma$ and 10\% for $\Delta\sigma_{z0}$. Center: efficiency of the fit 
of $\sigma$, $\Delta\sigma_{z0}$, $\Delta\sigma_{0x}$, $\Delta\sigma_{0y}$, 
$\Delta\sigma_{0z}$, $\Delta\sigma_{zx}$, $\Delta\sigma_{zy}$ and 
$\Delta\sigma_{zz}$ with the experimental $\phi$ distributions being smeared
by 3.5\% for $\sigma$ and 10\% for all the differences of cross sections. Right: same as center but
with the differences of cross sections being smeared by 5\%. 
The dotted histograms correspond to a reconstruction efficiency within 5\% , the dashed ones 
within 15\% and the solid ones within 25\%.}
\label{fig:effexp}
\end{figure}

For the open star topology,
the very small $\sigma_{(a)}$ values that one can observe for the $a(Im(H))$, $a(Im(\tilde{H}))$,
$a(Re(\tilde{E}))$ multipliers, compared to the $a(Re(H))$, $a(Re(E))$, $a(Re(\tilde{H}))$ 
and $a(Im(E))$ multipliers reflect their very high reconstruction efficiency and reliability.
This is confirmed by fig.~\ref{fig:effexp} which shows for the 
open-star topology, the reconstruction efficiencies
of each GPD multiplier. The three GPD multipliers just outlined have clearly
the highest reconstruction efficiencies. Comparing the open stars of fig.~\ref{fig:effexp} to 
the solid stars of fig.~\ref{fig:eff}, we note a loss of reconstruction 
efficiency when realistic simulated data are fitted, which is 
particularly important for $Im(E)$ and $Re(E)$. For these two GPD multipliers,
the reconstruction efficiency is smaller by a factor of 2 or more. 

In Fig.~\ref{fig:sig}, the open circle symbols (i.e. the fit to only 
$\frac{\Delta\sigma_{z0}}{\sigma}$) show up essentially
only in the $Re(H)$ and $Im(H)$ plots. However, the difference with the
solid-circle topology (i.e. the fit to $\sigma$ and $\Delta\sigma_{z0}$ separately)
is that most of the points are now NOT within the three $\sigma_{a()}$ bands, in 
spite of the relatively small $\sigma_{a()}$ values. This can lead to dangerous
interpretations where we might believe that the fitted value of, say $Im(H)$,
are extracted with high precision because of the small $\sigma_{a()}$ while the
value is not necessarily reliable. Firstly, this suggests that a detailed and dedicated error
analysis must be done when dealing with underdetermined systems such as this one 
where one fits only one observable by seven parameters. We will detail this in the next section. 
Secondly, let us emphasize that this in no way means that BSAs are dangerous
or not useful, but that either they should be complemented by some other observables
that provide additional constraints or that some physics input (such as the dominance 
of $H$ over the other GPDs) must guide the fitting procedure.
We recall that we have blinded ourselves, in this study, by not telling the fit 
which values it should start with. 

Let us also note in fig.~\ref{fig:events} that the error
bars of the open star topology are always smaller
than the ones from the open squares, which is consistent with the fact
that fitting eight observables rather than six improves the quality of the fit. 
Making a blunt generalization from this particular event
one can also conclude that it is almost equivalent to fit eight observables 
with a precision of 10\% (left, open stars) as to fit six observables 
with a precision of 5\% (right, open squares).
  
In this section, we have simulated real experimental conditions by (Gaussian-)smearing,
with a standard deviation equal to realistic experimental error bars, the 
observables that we fit with our code. We conclude that the fitting procedure is, 
though somewhat less efficient, still 
very reliable (the $E$ CFFs suffering the most important loss of reconstruction 
efficiency) and that the general conclusions that we drew in section~\ref{acad} remain 
valid. 
Having gained confidence in our code and fitting procedure through these simulations
with pseudo-data, we now proceed in the next section by applying the code to some real data.

\section{Application}
\label{halla}

The JLab Hall A collaboration has recently measured $\sigma$ and 
$\Delta\sigma_{z0}$~\cite{franck} at $x_B$=0.36 and $Q^2$=2.3 GeV$^2$
for four different $t$ values. From the previous sections, 
we have learned that the simultaneous fit of these two observables allows us to 
access $Im(H)$ at a relatively high efficiency level and to a somewhat 
lesser extent a few other CFFs such as $Re(H)$. 

\begin{figure*}[htb]
\epsfxsize=13.cm
\epsfysize=13.cm
\epsffile{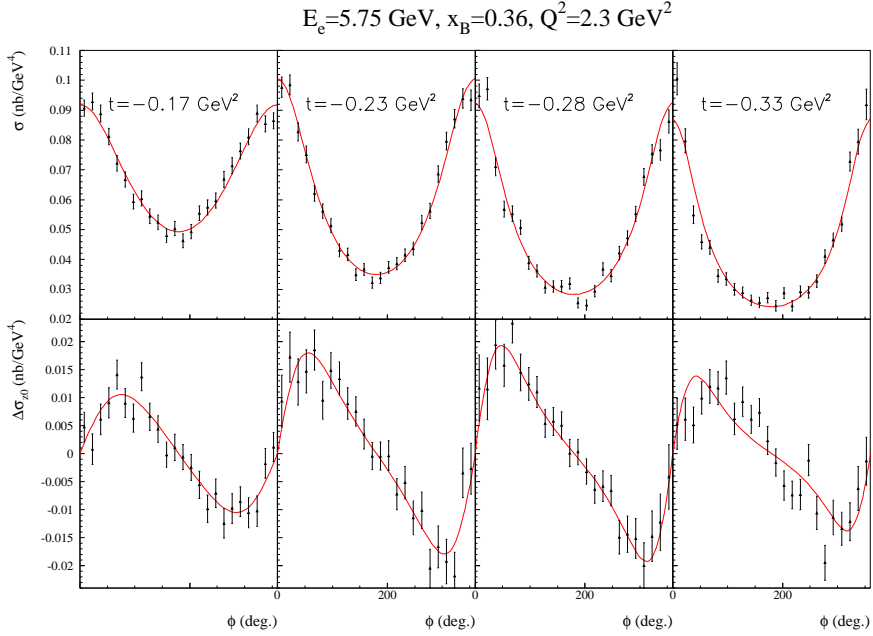}
\vspace{-3.cm}
\caption{Top row: DVCS $\sigma$, and bottom row: DVCS $\Delta\sigma_{z0}$ 
as a function of $\phi$, at $x_B$=0.36 and $Q^2$=2.3 GeV$^2$ 
for four different $t$ values. The data are from the JLab Hall A 
collaboration~\cite{franck}. 
The curves are the results of the fit from our code to the
$\sigma$ and $\Delta\sigma_{z0}$ observables. The fitted 
GPD multipliers which produce these curves are displayed 
in table~\ref{tab:param2}.}
\label{fig:halla}
\end{figure*}

We have therefore run our fitting code using the JLab Hall A data. 
Fig.~\ref{fig:halla} shows our results. The numerical results of the
fit for each GPD multiplier are displayed in table~\ref{tab:param2}.

\begin{table*}[htb]	
\begin{center}	
\begin{tabular}{|c|c|c|c|c|c|c|c|c|c|}
\cline{3-10}
\multicolumn{2}{c|}{}
&$Re(H)$  & $Re(E)$ & $Re(\tilde{H})$ & $Re(\tilde{E})$ & $Im(H)$ & $Im(E)$ &
$Im(\tilde{H})$& $\chi^2/N_{dof}$\\
\hline t=-0.17 GeV$^2$&$a()$ & -5.00 & -5.00 & -2.58 & -0.55 & 1.11 & -4.57 & -0.44 & 1.01\\
\cline{2-9}
& $\sigma^-_{a()}$  & $\infty$ & $\infty$ & -1.17 & -2.45 & -0.94 & $\infty$ & -2.71 & \\
\cline{2-9}
& $\sigma^+_{a()}$  & $\infty$ & $\infty$ & $\infty$ & $\infty$ & 0.18 & $\infty$ & $\infty$ & \\
\hline
t=-0.23 GeV$^2$&$a()$ & 0.22 & -5.00 & -5.00 & -2.07 & 1.17 & -2.61 & -0.89 & 0.92\\
\cline{2-9}
&$\sigma^-_{a()}$  & -1.03 & $\infty$ & $\infty$ & -1.55 & -0.98 & $\infty$ & -2.46 & \\
\cline{2-9}
&$\sigma^+_{a()}$  & 4.52 & $\infty$ & $\infty$ & 4.02 & 0.17 & $\infty$ & $\infty$ & \\
\hline
t=-0.28 GeV$^2$&$a()$ & 1.64 & -5.00 & -5.00 & -0.92 & 1.27 & -1.15 & -1.26 & 1.44\\
\cline{2-9}
&$\sigma^-_{a()}$  & -0.81 & $\infty$ & $\infty$ & -2.52 & -0.97 & $\infty$ & -1.75 &  \\
\cline{2-9}
&$\sigma^+_{a()}$  & 2.66 & $\infty$ & $\infty$ & 1.92 & 0.10 & $\infty$ & $\infty$ & \\
\hline
t=-0.33 GeV$^2$&$a()$ & 3.82 & 5.00 & 5.00 & 0.87 & 1.26 & -4.00 & -2.36 & 2.31\\
\cline{2-9}
&$\sigma^-_{a()}$  & -0.64 & $\infty$ & $-2.81$ & -2.36 & -0.28 & $\infty$ & -1.5 & \\
\cline{2-9}
&$\sigma^+_{a()}$  & 0.63 & $\infty$ & $\infty$ & 2.36 & 0.04 & $\infty$ & 2.18 & \\
\hline
\end{tabular}	
\caption{Fitted GPD multipliers $a()$ and their negative ($\sigma^-_{a()}$) and positive
($\sigma^+_{a()}$) uncertainties resulting from the fit of the JLab Hall A $\sigma$ and 
$\Delta\sigma_{z0}$ observables. The $\chi^2/N_{dof}$ value for these fits is presented
in the rightmost column. As explained in the text, only central values with finite
negative and positive uncertainties have a reliable meaning.} 
\label{tab:param2}
\end{center}	
\end{table*}	

Several issues are to be discussed concerning this table. Let us start with the uncertainties
on the fitted parameters. In the previous section, the uncertaintites 
quoted were the standard quadratic errors from MINUIT, i.e. estimated
from the second derivative of the $\chi^2$ function to be minimized with respect to the
parameter under consideration, based on the assumption that the $\chi^2$ function 
is parabolic near its minimum. This was sufficient for our general purpose of 
statistically estimating global reconstruction efficiencies and general features of our code. Now, 
in the present case with real data, the determination of the uncertainty on 
the fitted parameters has to be refined. Indeed, we are facing a multidimensional 
problem (seven parameters to fit) which 1) is non-linear, 2) has potentially strong correlations 
between the parameters and 3) is severely underconstrained with only two observables
to fit seven CFFs. Therefore, a careful and detailed error analysis 
was carried out with the MINOS subpackage of MINUIT which allows us to
explore in a gradual and automated way the $\chi^2$ landscape around the minimum
and define one-standard deviation uncertainties for each parameter when it reaches $\chi^2+1$.
This method, which is essential for non-linear problems, yields 
asymmetric error bars. These are presented with the symbols $\sigma^-_{a()}$ and 
$\sigma^+_{a()}$ (the negative and positive errors, respectively) in table~\ref{tab:param2}.
The price one pays is that MINOS is very time-consuming and that the error bar determination
of our seven parameters took several hours of computing time. 

In table~\ref{tab:param2}, the symbol $\infty$ means that the uncertainty coming out of MINOS 
is very large, i.e. exceeding the limits given in MINOS. To be precise, the results 
in table~\ref{tab:param2} have been obtained by setting, like in the previous section,
the lower and upper limits of the domain of variation allowed for the seven parameters 
to -5 and 5 respectively. The $\infty$ symbol means that $\chi^2+1$ was never reached
in the interval considered. It can be interpreted as a whole range of 
values for the associated parameter can accomodate the fit with relatively equally 
good $\chi^2$. 
In other words, the corresponding GPD multiplier (or equivalently CFF) is essentially 
unconstrained and therefore no 
particular confidence and meaning can be given to the quoted numerical value. This does not 
mean that the values associated with an $\infty$ uncertainty in table~\ref{tab:param2}
can take any value, irrespective of the other multipliers. Many of the seven parameters are 
highly correlated with other parameters and changing the value of one parameter 
will change the values of others. This means that for those parameters with an 
$\infty$ uncertainty, the solution is not unique and that other sets of
values are also possible. For instance, fixing some GPD multipliers to 
particular values 0, or 1 (the associated CFF takes its
VGG value in this latter case) 
and letting all other parameters remain free can result in fits with an almost 
equivalent $\chi^2$ compared to letting all seven parameters free.

However, what is remarkable is that even though many CFFs have an $\infty$ uncertainty
in table~\ref{tab:param2}, a few of them, depending on their $t$ values, come out
with finite error bars. For instance, a central value with finite negative and positive
uncertainties can be extracted for $a(Im(H))$ at all $t$ values and for $a(Re(H))$ 
at the largest three $t$ values. For these multipliers, the uncertainties, while finite, 
can be very large, 
the negative errors ranging from more than 100\% at the smallest $t$ values to $\approx$
20\% at the largest $t$ values. This should not come as a surprise, given
the complexity and the underdetermination of our problem.
Therefore, the large uncertainties that we obtain 
don't reflect a lack of quality or precision of the experimental data 
but rather a lack of constraints and of sensitivity of the observables to certain parameters. 
In these conditions, it is already a significant success to be able to pull out 
some reliable and stable information from such partial inputs. We restate that in this
study we have left all seven GPD multipliers (or equivalently CFFs) free. If some of them can be 
fixed, neglected or constrained to
a domain smaller than $\{-5,5\}$, it is clear that these uncertainties can drastically
be reduced. This will be the subject of another forthcoming article. We recall
that when enough experimental 
observables are available to fit,
as we have shown in the previous sections with simulations, such assumptions won't be
necessary and all parameters should be uniquely determined.

We have carried out many checks to get confidence in the numbers (central
values and error bars) associated with finite uncertainties in 
table~\ref{tab:param2}. For instance, to start the minimization, MINUIT requires us to 
set some starting values for the parameters to be fitted. As we mentioned in the previous 
sections, our strategy is in general to set them to the VGG values in order to be, hopefully,
close to the true solution. However, we checked that by starting with randomly generated starting 
values, MINUIT+MINOS were always converging towards the same central values 
given in table~\ref{tab:param2} and producing the same corresponding uncertainties at the few 
percent level. 

Another check was to vary the limits of the domain of variation of the 
parameters (initially set to $\{-5,5\}$). We made the fits with
the domains $\{-3,3\}$ and $\{-7,7\}$. The outcome was that,
although all central values associated with $\infty$ uncertainties in table~\ref{tab:param2}
could be different, the few values associated with finite positive and negative uncertainties 
(i.e. $a(Im(H))$ and $a(Re(H))$) were found remarkably stable at the few percent level.
It was interesting to note that for the two intermediate $t$ values $a(Re(E))$ and 
$a(Re(\tilde{H}))$ would systematically take the lower limit of the domain (i.e. -3 or 
-7, similarly to the -5 in table~\ref{tab:param2}) though this had no effect on
the central values of $a(Im(H))$ and $a(Re(H))$. However,
a (moderate) effect was found for the uncertainties of $a(Im(H))$ and $a(Re(H))$. 
It was observed 
that their uncertainties, calculated by MINOS, were slowly increasing as the limits
of the domain of variation increased. This can certainly be attributed to the fact 
that the correlations between the ``stable" $a(Im(H))$ and $a(Re(H))$ multipliers
on the one hand and the ``unstable" $a(Re(E))$ and $a(Re(\tilde{H}))$ are not 
completely absent, and that larger (absolute) values of the latter multipliers (since they 
actually reach the limits of the domain of variation as mentionned earlier) can certainly
tend to increase the error bars of the former ones. A few paragraphs below, we make a quantitative
comparison of this (moderate) effect.

Although the uncertainties on the $a(Im(H))$ and $a(Re(H))$ are rather large,
we can notice a few trends. $a(Im(H))$ is always consistent with 1 within error bars
and it tends to increase with $t$. Although the purpose of this study is not at all
to prove (or disprove) the VGG model, it shows that the VGG model seems
to provide a reasonable estimate of $Im(H)$, possibly underestimated at the
$\approx$ 20\% level. The parameter $a(Re(H))$ has a strong tendancy to rise with $t$
and although the uncertainties are large, it seems to be significantly different
from the VGG prediction and it can differ by up to a factor of 4.


\begin{table*}[htb]	
\begin{center}	
\begin{tabular}{|c|c|c|c|c|c|c|c|}
\cline{2-8}
\multicolumn{1}{c|}{} &$Re(H)$  & $Re(E)$ & $Re(\tilde{H})$ & $Re(\tilde{E})$ & $Im(H)$ & 
$Im(E)$ & $Im(\tilde{H})$ \\
\hline 
t=-0.17 GeV$^2$ & 0.25 & 0.47 & 0.55 & 37.71 & 1.96 & 0.62 & 0.51 \\
\hline
t=-0.23 GeV$^2$ & 0.49 & 0.40 & 0.50 & 26.91 & 1.74 & 0.54 & 0.46 \\
\hline
t=-0.28 GeV$^2$ & 0.66 & 0.35 & 0.46 & 21.71 & 1.58 & 0.48 & 0.43 \\
\hline
t=-0.33 GeV$^2$ & 0.78 & 0.31 & 0.42 & 18.16 & 1.44 & 0.43 & 0.40 \\
\hline
\end{tabular}	
\caption{The reference VGG CFFs at $\xi$= 0.22 for the four JLab Hall A $t$ values. 
The multiplication of these reference VGG values by the fitted GPD multipliers 
of table~\ref{tab:param2} (and their associated error) yield the measured CFFs.} 
\label{tab:gpd}
\end{center}	
\end{table*}	

\begin{figure}[h]
\epsfxsize=10.cm
\epsfysize=10.cm
\epsffile{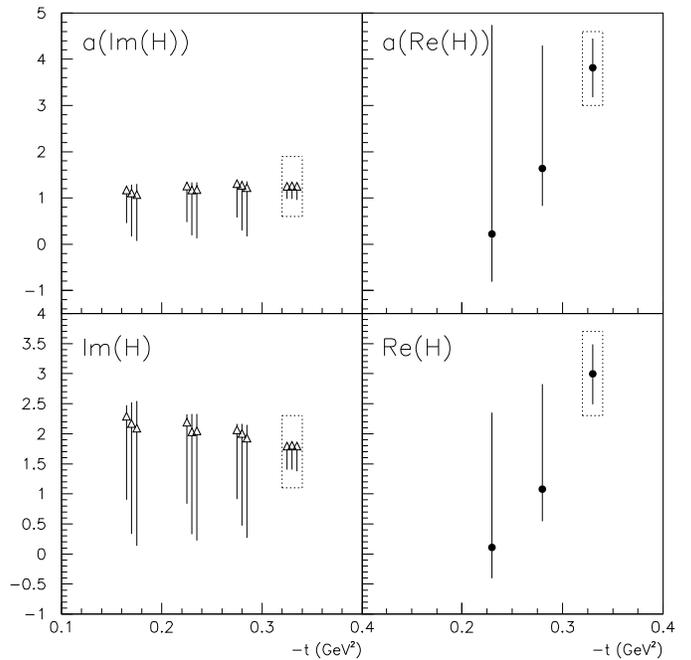}
\caption{Upper plots: the $a(Im(H))$ (left) and $a(Re(H))$(right) GPD multipliers 
as a function of $t$ extracted from the fits of the JLab Hall A data (fig.~\ref{fig:halla})
at $\xi$= 0.22. Lower plots: the resulting $Im(H))$ (left) and $Re(H)$(right) CFF values.
For $a(Im(H))$ and $Im(H)$, three results for each $t$ are presented corresponding
to fits with the domains of variations $\{-3,3\}$, $\{-5,5\}$ and $\{-7,7\}$ from left to right 
respectively. The dotted boxes around the largest $t$ results indicate that the 
associated fit has a $\chi^2$ about a factor 2 worse (see table~\ref{tab:param2}) than the fits 
at the other $t$ values.}
\label{fig:fin}
\end{figure}

We display in table~\ref{tab:gpd} the VGG reference values which,
by multiplying the $a()$ GPD multipliers and their associated errors of table~\ref{tab:param2}, 
allow us to extract the measured values of the CFFs themselves. We plot in fig.~\ref{fig:fin}
the measurements of the two multipliers $a(Im(H))$ and $a(Re(H))$ and the associated 
CFFs for the $t$ values for which they have finite uncertainties. In this figure,
we show three results for each $t$ for $a(Im(H))$ and $Im(H)$. These correspond
to the fits with the domains of variations $\{-3,3\}$, $\{-5,5\}$
and $\{-7,7\}$ from left to right, respectively. The 
stability of the central value as well as the moderate increase of the error bar
with the increase of the limit of the domain previously mentioned can be observed. 
As deviations of CFF
values by a factor 7 (or even 5) from VGG predictions are not easy to conceive, 
the uncertaintities quoted in table~\ref{tab:param2} and displayed in fig.~\ref{fig:fin} 
are very conservative. They could be reduced 
if, based on some models or educated guesses, smaller limits could be set 
on the other CFFs.

One should also note from table~\ref{tab:param2} that all fits have a $\chi^2$ close 
to 1 except for
the bin at $t=-0.33$ GeV$^2$. We remark that a problem with this same bin was
also observed in ref.~\cite{fxthese} where, in an analysis mainly devoted to
the JLab Hall B BSAs, the JLab Hall A BSAs, i.e. $\frac{\Delta\sigma_{z0}}{\sigma}$, were
fitted by the phenomenological function $\frac{\alpha\sin\phi}{1+\beta\cos\phi}$
where the $\beta$ coefficient was found to take a peculiar value for this
particular $t$ bin, in clear distinction from the surrounding $t$ bins.
Also, in the present analysis, although the results for $a(Re(E))$ and $a(Re(\tilde{H}))$ 
are not very meaningful because of their $\infty$ uncertainties, one can nevertheless 
clearly observe a discontinuity in their fitted values as they jump rather abruptly from
a largely negative values at the three low $t$ values to a largely positive value
for the last $t$. Therefore, there definitely seems to be a change of behavior in the data at this 
large $t$ bin. A bad $\chi^2$ may be indicating that our theoretical modeling
is not appropriate. We recall that our analysis has been carried out at the twist-2 handbag 
level and that this intriguing change might signal the rise of higher twist corrections.

Let us also mention that still a few more pieces of information can be extracted
from the values of table~\ref{tab:param2}. For several $t$ values, $a(Re(\tilde{E}))$ 
has finite uncertainties. Since these error bars are systematically over 100\%,
it makes the information difficult to use, but we can simply note that the
value of $1$, which corresponds to the VGG pion pole, is never excluded
and that our original assumption to neglect $Im(\tilde{E})$ on this basis
is therefore not ruled out. There are a few cases in table~\ref{tab:param2}
where only one uncertainty ($\sigma^-_{a()}$ in general) is finite, which
still yields some information. In particular, $Im(\tilde{H})$ 
for all $t$'s has a reliable (lower) bound.

We have also tried to fit only the $\Delta\sigma_{z0}$
observables (i.e. without the simultaneous fit of $\sigma$)
which would be sensitive only to $Im(H)$, $Im(E)$ and $Im(\tilde{H})$. 
In general, the fitted central values for these CFFs were unstable, unless
we guided and constrained the fit by setting limits on the 
domain of variation of the parameters. 
In general, we obtained values of $Im(H)$ about 30\% higher than those 
quoted in table~\ref{tab:param2} but they corresponded to much larger values
of $Im(\tilde{H})$, compared to those of table~\ref{tab:param2}, 
which with its opposite sign would provide some compensation.
This kind of unstability is not unexpected because fitting only one observable
with three parameters is even less constraining than fitting two
observables with seven parameters. 

Finally, we tried to fit the numerous JLab
Hall B BSAs~\cite{fx}. We found that without any guidance of the fit or 
any physics input, no fit stability and robustness could be reached. 
This could actually be anticipated from our studies in the previous section.
However, as we saw with fig.~\ref{fig:events}, guiding the fit by setting
the initial values in MINUIT to values close to the expected ones,
which can be relatively safely done for $Im(H)$ following our results
on the Hall A data that we just presented, could tremendously enhance the fit efficiency
and stability. We delay to a forthcoming publication the results of such guided 
or educated fits.

\section{Conclusion}

In sumary, we have developed a fitter program 
to extract GPD information from various DVCS observables
at QCD leading twist and leading order.
We have gained confidence in the program by checking its 
reliability and efficiency on simulated, ideal or realistic, pseudo-data.
We came to the conclusion that by fitting enough unpolarized, singly and doubly 
polarized observables, with realistic experimental error bars, the code
was able to extract in most cases the seven CFFs considered in this work, even in a very blind
framework where all parameters are left free and essentially unconstrained.
Applying realistic and educated constrains, such as dispersion relations
or model motivated ansatzes, can only reduce the number of independent 
parameters or limit the range of variation of certain parameters
and therefore improve the efficiency and reliability of this 
fitter program. We recall that most of the aforementionned observables will be available 
in the near future from various experiments at JLab in particular. If only some of these 
nine observables are available, 
we have demonstrated that valuable partial GPD information could still be extracted:
in particular, with $\sigma$, $\Delta\sigma_{z0}$ and
$\Delta\sigma_{0z}$, which are planned to be measured in 
the very near future, the $Im(H)$, $Im(E)$ and $Im(\tilde{H})$ CFFs could
be very reliably be known. For the real part CFFs, one can either
measure the beam charge difference of cross sections or 
double polarization observables. 


Finally, we have extracted from the 
JLab Hall A data first numerical values with associated uncertainties for the 
$H$ GPD combinations $H(\xi , \xi, t) - H(- \xi, \xi, t)$ and 
$P \int_0^1 d x \left[ H(x, \xi, t) - H(-x, \xi, t) \right] C^+(x, \xi)$.
The corresponding error bars are rather large because we are fitting more
parameters than observables and no a priori knowledge of any CFFs has been assumed in 
this study.
Ultimately, when enough observables are available, this should indeed not be needed,
and we can consider that we have placed ourselves in this long-term perspective. 
In a forthcoming article, we will discuss strategies to reduce these uncertainties 
on the present data if educated assumptions are made about the CFFs.
This code is only the first step towards a general fitting procedure
of DVCS data (and potentially, if data lend themselves to a GPD interpretation,
of exclusive meson electroproduction). Numerous extensions of the code
are possible, such as introducing QCD higher twists and/or higher order corrections, 
implementing dispersion relations, and so forth.

The author has benefitted from numerous discussions with many colleagues 
from JLab and from the French ``Nucleon GDR" group, whose list would be too 
long to enumerate but to whom the author is very thankful. 
Very special thanks are given to P. Desesquelles, M. Gar\c{c}on, K. Griffioen,
C. Mu\~noz Camacho and M. Vanderhaeghen.


\begin{footnotesize}

\end{footnotesize}

\end{document}